\documentclass[conference,a4paper]{IEEEtran}
\IEEEoverridecommandlockouts

\makeatletter
\def\markboth#1#2{\def\leftmark{\@IEEEcompsoconly{\sffamily}\MakeUppercase{\protect#1}}%
\def\rightmark{\@IEEEcompsoconly{\sffamily}\MakeUppercase{\protect#2}}}
\makeatother

\usepackage[english]{babel}
\usepackage{color}
\usepackage{lipsum}% http://ctan.org/pkg/lipsum
\usepackage{caption}
\usepackage{cite}
\usepackage[pdftex]{graphicx}
\usepackage{subcaption}
\usepackage{amsmath}
\usepackage{amsfonts}
\usepackage{array}
\usepackage{verbatim}
\usepackage{listings}
\usepackage{algorithm}
\usepackage{algpseudocode}
\usepackage{hyperref}
\usepackage{url}
\usepackage{enumerate}
\usepackage{multirow}

\usepackage{epsfig}
\usepackage{epstopdf}
\usepackage{multicol}% http://ctan.org/pkg/multicols
\usepackage[font=footnotesize]{caption}

\renewcommand{\arraystretch}{2}

\newcommand{\bi}{\begin{itemize}}
\newcommand{\ei}{\end{itemize}}
\newcommand{\be}{\begin{equation}}
\newcommand{\ee}{\end{equation}}

\def\beq{\begin{equation}}
\def\eeq{\end{equation}}
\def\beqa{\begin{eqnarray}}
\def\eeqa{\end{eqnarray}}
\def\beqan{\begin{eqnarray*}}
\def\eeqan{\end{eqnarray*}}

\usepackage{threeparttable}
\usepackage[table,xcdraw]{xcolor}
\usepackage{tabularx}
   \usepackage{multirow}
   \usepackage{booktabs}

   \usepackage{graphicx}
   \usepackage{array, blindtext}
   \usepackage{wrapfig}
\usepackage{pdfpages}

\title{\mbox{ \hspace{-0.3899 cm}Multi-Connectivity \hspace{-0.4 cm} in  \hspace{-0.4 cm} 5G  \hspace{-0.4 cm} mmWave  \hspace{-0.4 cm} Cellular  \hspace{-0.4 cm} Networks}}
\author{{{\bf Marco Giordani}$^\dagger$, {\bf Marco Mezzavilla}$^\diamond$, {\bf Sundeep Rangan}$^\diamond$, {\bf Michele Zorzi}$^\dagger$ }\\
$^\dagger$ University of Padova, Italy \quad $^\diamond$NYU Wireless, Brooklyn, NY, USA \\
emails: \small{$\{$\texttt{giordani}, \texttt{zorzi}$\}$\texttt{@dei.unipd.it}, $\{$\texttt{mezzavilla}, \texttt{srangan}$\}$\texttt{@nyu.edu}
}}

\begin{document}
\maketitle

\begin{abstract}  The millimeter wave (mmWave) frequencies offer the potential of orders of magnitude
increases in capacity for next-generation cellular wireless systems.  However, links in mmWave networks are highly
susceptible to blocking and may suffer from rapid variations in quality.
Connectivity to multiple cells -- both in the mmWave and  in the traditional lower frequencies -- is thus considered
essential for robust connectivity. However, one of the challenges in supporting multi-connectivity in the mmWave
space is the requirement for the network  to track the direction of each link in addition to its
power and timing.  With highly directional beams and fast varying channels, this directional tracking
may be the main bottleneck in realizing robust mmWave networks.
To address this challenge, this paper
proposes a novel measurement system based on (i) the UE transmitting sounding signals in
directions that sweep the angular space, (ii) the
mmWave cells measuring the instantaneous received signal strength along with its variance to better capture the dynamics and, consequently, the reliability of a channel/direction and, finally, 
(iii) a centralized controller making handover and scheduling decisions based on the mmWave cell
reports and  transmitting the decisions either via a mmWave cell or conventional microwave cell (when control signaling paths are not available).  
We argue that the proposed scheme enables efficient and highly adaptive 
cell selection in the presence of the channel variability  expected at mmWave frequencies.
\end{abstract}


\section{Introduction}

\begin{figure*}[t!]
\centering
 \includegraphics[trim= 0cm 0cm 0cm 0cm , clip=true, width= \textwidth]{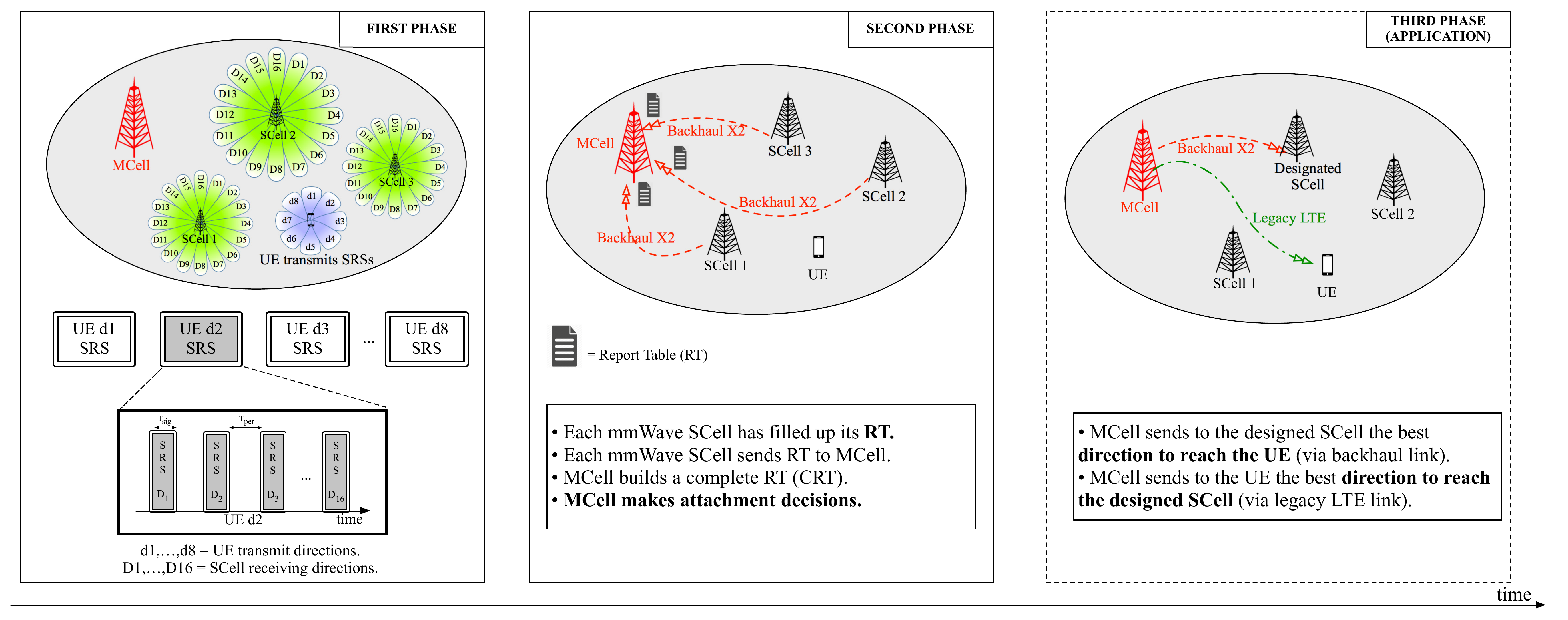}
 \caption{Slot scheme for the proposed MC procedure. After the first phase, each mmWave SCell builds a report table which is used to track the propagation conditions of its surrounding channels. Green and red dashed lines refer to the control messages exchanged  via the legacy communication link and the backhaul X2  connection, respectively. In this figure, we assume that the MCell is identified as the macrowave base station, which performs the network monitoring. In this example, $N_{\rm BS} = 16$ and $N_{\rm UE} = 8$.}
 \label{fig:slot_scheme}
\end{figure*}

The millimeter wave (mmWave) bands -- roughly above 10~GHz --
have attracted considerable attention
for micro- and picocellular systems \cite{RanRapE:14}.
These frequencies offer much more
bandwidth than current cellular allocations in the congested bands below 3~GHz, and initial capacity estimates have suggested that networks with
mmWave cells can offer orders of magnitude greater capacity than state-of-the-art 4G systems \cite{Mustafa}.

However, one of the challenges in designing cellular networks in the mmWave bands is robustness,
and one likely key feature of mmWave cellular networks that can improve robustness is
\emph{ multi-connectivity} (MC) \cite{MC}.
MmWave signals are blocked by many common building materials such as brick, and  the human body can also
significantly attenuate signals in the mmWave range \cite{lu2012modeling}.  Thus, the communication quality
between the user equipment (UE)
and any one cell can be highly variable as the movement of obstacles or even the changing position of the
body relative to the mobile device can lead to rapid drops in signal strength.
Multi-connectivity is a feature in which each UE maintains multiple possible signal paths to different
cells so that drops in one link can be overcome by switching data paths.
In  mmWave networks, this multi-connectivity can be  both among multiple mmWave cells
as well as between 5G mmWave cells and traditional 4G cells below 3~GHz, with the low frequency cells offering greater
robustness but lower bandwidth.

This paper addresses one of the key challenges in supporting multi-connectivity in
heterogeneous networks (HetNets) with mmWave cells, namely directional multi-cell channel tracking
and measurement reports.
Multi-connectivity is already supported as part of carrier aggregation -- one of the most
important features in 3GPP LTE-Advanced \cite{CA_LTE}.  In standard carrier aggregation,
the network maintains power and timing information of the UE at multiple cells and can then
schedule data on the optimal link.
However, multi-connectivity is significantly more complicated at mmWave, primarily because
transmissions are likely to be highly directional. Thus,
in addition to power and timing tracking, the network and UE must constantly monitor the direction
of transmission of each potential link. Tracking changing directions can slow the rate at which
the network can adapt and can be a major obstacle in providing robust service in the face of variable link
quality.  In addition, the UE and the base station (BS)  may only be able to listen to one direction at a time,
thus making it hard to receive the control signaling necessary to switch paths.

To address this challenge, this paper proposes a novel multi-cell
measurement reporting system 
where each UE
 directionally broadcasts a sounding reference signal (SRS) in a time-varying direction that continuously
 sweeps the angular space. Each potential serving cell scans all its angular directions
 and monitors the strength of the received SRS along with its variance, to better capture the dynamics of the channel. A centralized controller obtains complete directional knowledge from all the potential cells
 in the network to make the optimal serving cell selection and
 scheduling decision.  We argue that this scheme has the three important features listed below.

\textbf{Directional uplink measurements:} Importantly, unlike in
traditional LTE channel aggregation, the proposed system is based on
the channel quality of uplink (UL) rather than downlink (DL) signals. We argue that this has several key benefits. First, 
the use of UL signals eliminates the need for the UE
to send measurement reports back to the network and thereby removes a point of failure in the control
signaling path.
Second, we show that if digital beamforming or beamforming with multiple analog streams
is available at the mmWave cell, then the
directional scan time can be dramatically reduced when using UL-based
measurements.  
Finally, since the base station is less power constrained
than a mobile device, digital or hybrid beamforming will likely be more feasible
at the BS side.

\textbf{Capturing channel variance:} Additionally, by updating the variance of the per user received signal strength at each iteration, we are able to better capture the dynamics of the channel and bias the cell selection strategy of delay-sensitive applications towards more robust cells.

\textbf{Enhanced control signaling:} A key issue in implementing control for
mmWave multi-connectivity is directionality.  Specifically, if the UE is using
analog beamforming it may only be able to ``listen" in one direction at a time.
Hence, if the network switches the serving cell (handover), the new serving cell may not be able
to communicate the switch to the UE.  To circumvent this problem, we propose
that the network be able to send scheduling and serving cell decisions of the mmWave cells
over the legacy microwave cells, in the case when the mmWave links are not available.
Control signaling can be used to implement more efficient resource allocation on initial access strategies.

The rest of the paper is organized as follows. In Section \ref{sec:MCP}, we describe the multi-cell measurement procedure. In Section \ref{sec:hints}, we report some examples of how this framework can be used towards designing better control strategies. In Section \ref{sec:res}, we provide some results and, finally,  we conclude the paper and list some future research steps in Section \ref{conclu}.

%In addition to the uplink signaling, the paper also discusses methods for sending
%control signal back to the UE over 4G channels and for maintaining statistical information on the
%channel quality to track variability.

%The MiWEBA system concept rationale is to overcome the current limitations by an integrated holistic approach using mm-wave technology. A fundamental ingredient is the separation of control and user-plane. The control plane (C-plane) will be operated at the regular cellular frequencies e.g. between $700$ MHz and $2.7$ GHz and the data link layers (U-plane) of the cellular network and the millimetre-wave band as a high capacity deployment overlay. In this overlay the wide coverage area of the c-plane allows to track user links and exchange information on channel state and link quality and therefore to enable seamless handover between the robust low and moderate data rate cellular band and the high data rate mm-wave band.

%\newpage

%\begin{figure}[t!]
%\centering
% \includegraphics[trim= 0cm 0cm 0cm 0cm , clip=true, width=0.7 \columnwidth]{Draws/BS_UE}
% \caption{BS (left) and UE (right) beam pattern. . Each BS is equipped with a $8 \times 8$ arraty and can steer beams in $N_{\rm BS}=16$ directions, to cover the whole $360^\circ$ angular space. The user exploits an array of $4 \times 4$ antennas, steering beams through $N_{\rm UE}=8$ angular directions. }
% \label{fig:BS_UE}
%\end{figure}

\section{Procedure Description}
\label{sec:MCP}

%\subsection{System Model}

In the proposed framework, illustrated in Figure \ref{fig:slot_scheme}, there is one major node called MCell (Master Cell, in accordance with  3GPP LTE terminology), which here is typically
a microwave base station. However, functionally, the MCell can be any network entity
that performs centralized handover and scheduling decisions.  
The UE may receive data from a number of cells, either mmWave or microwave,
 and we call each such cell an SCell (Secondary Cell).
In order to communicate and exchange control information, the SCells and the MCell are inter-connected via traditional backhaul X2 interface connections, while each user can be reached by its serving MCell through the legacy 4G-LTE band.

MmWave SCells and UE will likely utilize directional phase arrays for transmission.
In this work, we will assume that nodes select one of a finite number of directions, 
and we let $N_{\rm BS}$ and $N_{\rm UE}$ be the number of directions at each BS 
and UE, respectively.  Thus, between any cell and the UE there are a total of 
$N_{\rm BS} \times N_{\rm UE}$ direction pairs.  The key challenge in implementing 
multi-cell connectivity is that the network must, in essence, monitor the signal strength 
on each of the directions pairs for each of the possible links. This is done by each SCell  building a report tables (RT), based on the channel quality of each receiving direction, per each user.  Our proposed method
performs this monitoring through the following three phases.

\subsubsection{Phase 1:  Uplink measurements}

The UE directionally broadcasts uplink sounding reference signals in dedicated slots, steering through directions $1,\dots,N_{\rm UE}$, one  at a time, to cover the whole angular space.
The SRSs are scrambled by locally unique identifiers (e.g., C-RNTI) that are known to the 
SCells. Each SCell performs an exhaustive search, scanning through $N_{\rm BS}$ directions,\footnote{In the case of digital beamforming, the receiver would detect the signal strength from all directions in a single slot.} in order to fill  the $i^{\rm th}$ row of the report table, which refers to the user steering direction $i$. The quantity

\begin{equation}
\text{SINR}_{i,j} = \max_{k=1,\dots,N_{\rm BS}} \text{SINR}_{i,j}(k)
\label{eq:max_SINR}
\end{equation}
represents the highest perceived SINR between the UE, transmitting through direction $i$, and SCell$_j$, maximized over all its possible receiving directions. The value
\begin{equation}
\text{d}_{i,j} = d(\text{SINR}_{i,j}) = d\Big(\max_{k=1,\dots,N_{\rm BS}}  \text{SINR}_{i,j}(k)\Big)
\end{equation}
is the angular direction through which such $\text{SINR}_{i,j}$ was received by SCell$_j$.

Each mmWave SCell 
keeps a record of previous RTs and updates, at each scan, the variance var$_{i,j}$ of the maximum SINR, $\text{SINR}_{i,j}$. When, at scan $t$,  SCell$_j$ computes a new SINR value $\text{SINR}_{i,j}^{(t)}$, according to \eqref{eq:max_SINR}, the variance is updated as:
\begin{equation}
\begin{aligned}
&\text{var}_{i,j}^{(t)} = \text{var} \Big(\text{SINR}_{i,j}^{(1)}, \dots, \text{SINR}_{i,j}^{(t)} \Big) \\ 
& =\frac{ \sum_{h=1}^t \Big(\text{SINR}_{i,j}^{(h)}\Big)^2 }{t} - \left(\frac{ \sum_{h=1}^t \text{SINR}_{i,j}^{{(h)} }}{t}\right)^2
\end{aligned}
\end{equation}
using the SINR values of the previously saved report tables.

If the base station has finite memory and can keep a record of just $U$ previous RT replicas, then the variance is updated as:
\begin{align}
&\text{var}_{i,j}^{(t)} = \text{var} \Big(\text{SINR}_{i,j}^{(t-U)}, \dots, \text{SINR}_{i,j}^{(U)} \Big) \\ 
& =\frac{ \sum_{h=t-U}^U \Big(\text{SINR}_{i,j}^{(h)}\Big)^2 }{U} - \left(\frac{ \sum_{h=t-U}^U \text{SINR}_{i,j}^{(h)} }{U}\right)^2 \notag
\end{align}

The uplink SRS signals could also be monitored by cells that are not currently SCells
to see if they should be added.\\

\begin{table}[!t]
\centering
\renewcommand{\arraystretch}{1.2}% Tighter
\begin{tabularx}{1\columnwidth}{@{\extracolsep{\fill}}c |c| c| c |c}
\toprule
\textbf{UE direction} & \textbf{SCell$_1$ } &\textbf{SCell$_2$ } & \textbf{\dots} & \textbf{SCell$_M$ }\\
\hline
\textbf{1} &\begin{tabular}{@{}c @{}} SINR$_{1,1}$ \\ $d_{1,1}$ \\ var$_{1,1}$ \end{tabular} & \begin{tabular}{@{}c @{}} SINR$_{1,2}$ \\ $d_{1,2}$ \\ var$_{1,2}$ \end{tabular} & \dots & \begin{tabular}{@{}c @{}} SINR$_{1,M}$ \\ $d_{1,M}$ \\ var$_{1,M}$ \end{tabular}\\
\hline
\textbf{2} &\begin{tabular}{@{}c @{}} SINR$_{2,1}$ \\ $d_{2,1}$ \\ var$_{2,1}$\end{tabular} & \begin{tabular}{@{}c @{}} SINR$_{2,2}$ \\ $d_{2,2}$\\ var$_{2,1}$ \end{tabular} & \dots & \begin{tabular}{@{}c @{}} SINR$_{2,M}$ \\ $d_{2,M}$ \\ var$_{2,M}$ \end{tabular}\\
\hline
\dots &\dots & \dots &\dots &\dots \\
\hline
\textbf{$N_{\rm UE}$} &\begin{tabular}{@{\extracolsep{\fill}}c} SINR$_{N_{\rm UE},1}$ \\   $d_{N_{\rm UE},1}$\\ var$_{N_{\rm UE},1}$ \end{tabular} & \begin{tabular}{@{}c @{}} SINR$_{N_{\rm UE},2}$ \\ $d_{N_{\rm UE},2}$\\ var$_{N_{\rm UE},2}$ \end{tabular} & \dots & \begin{tabular}{@{}c @{}} SINR$_{N_{\rm UE},N_{\rm UE}}$ \\ $d_{N_{\rm UE},N_{\rm UE}}$ \\ var$_{N_{\rm UE},N_{\rm UE}}$ \end{tabular}\\
\bottomrule
\end{tabularx}
\caption{An example of the complete report table that the MCell builds, after having received the partial RTs from the $M$ surrounding mmWave SCells in the considered area. We suppose that the UE can send the sounding signals through $N_{\rm UE}$ angular directions.}
\label{tab:RT}
\end{table}

\subsubsection{Phase 2:  Network decision}
Once the RT of each SCell has been filled, each mmWave cell sends 
this information to the supervising microwave MCell through the backhaul link which, in turn, builds a complete report table (CRT), as depicted in Table \ref{tab:RT}.
When accessing the CRT, the MCell  selects the best mmWave BS candidate for the considered user, based on different metrics. For example, the MCell could select the maximum SINR
(with some hysteresis), in order to have the best channel propagation conditions, so
\begin{equation}
( \text{CRT} \, \text{row}, \text{CRT} \, \text{col}) = (d_{\rm UE},n_{\rm ID}) = \max_{\substack{i=1,\dots,N_{\rm UE}\\ j =1,\dots,M}} {\rm SINR}_{i,j},
\end{equation}
where $d_{\rm UE}$ is the direction the UE should set to obtain the maximum SINR and reach the mmWave SCell with ID $n_{\rm ID}$.
Such maximum SINR is associated, in the CRT's entry, to the SCell direction $d_{\rm SCell} $, which should therefore be selected by the mmWave BS to reach the UE with the best  performance.

\subsubsection{Phase 3:  Path switch and scheduling command}
If the serving cell needs to be switched, or a secondary cell needs to be added or dropped,
 the MCell needs to inform both the UE and the cell.
Since the UE may not be listening in the direction of the target SCell,
the UE may not be able to hear a command from that cell.  Moreover, since a common reason for
the path switch and for cell additions in the mmWave regime 
are due to link failures, the control link to the serving mmWave cell may not be available either.
To handle these circumstances, we propose that the path switch and scheduling commands
be able to be communicated over the legacy 4G cell.  Therefore, the MCell notifies  the designated mmWave SCell (with ID $n_{\rm ID}$), via the high capacity backhaul,  about the UE's desire to attach to it. It also embeds the best direction $d_{\rm SCell}$ that should be set to reach that user. Moreover, it sends to the UE, through an omnidirectional control signal at microwaves, the best user direction $d_{\rm UE}$ to select, to reach such candidate SCell. By this time, the best SCell-UE beam pair has been determined, therefore the transceiver can directionally communicate in the mmWave band.

%
%At this time, we suppose that the UE has already set up a link to the macro cell base station, on a legacy LTE connection. The MCell feeds back, via the high capacity backhaul, to the designated mmWave SCell (with ID $n_{\rm ID}$) about the UE's desire to attach to it. It also embeds the  best  direction  $d_{\rm SCell}$ that should be set, to reach the user. Moreover, it sends to the UE, through an omnidirectional control signal at microwaves,	the best user direction $d_{\rm UE}$ to select, to reach such candidate SCell.
%By this time, the best SCell-UE beam pair has been determined, therefore the transceiver can directionally communicate in the mmWave band.
%

\section{Control Applications for MC Procedure}
\label{sec:hints}

The proposed UL-based framework can be used to address some of the most important 5G control plane challenges that arise when dealing with mmWave frequency bands. In this section, we list some of the functions that are suitable to be implemented with our MC procedure.

\subsection{Handover}
\label{sec:HO}

Handover  is performed when the UE moves from the coverage of one cell to the coverage of another cell \cite{LTE_book}. Frequent handover, even for fixed UEs, is a potential drawback of mmWave systems due to their vulnerability to random obstacles, which is not the case in LTE. Dense deployments of short range BSs, as foreseen in mmWave cellular networks, may exacerbate frequent handovers between adjacent BSs. Loss of beamforming information due to channel change is another reason for handover and reassociation \cite{zorzi}. There are only a few papers on handover in mmWave 5G cellular \cite{handoff,HO_train,HO_het,HO_het_2}, since research in this field is just in its infancy.

\begin{figure}[t!]
\centering
 \includegraphics[trim= 0cm 0cm 0cm 0cm , clip=true, width=0.9 \columnwidth]{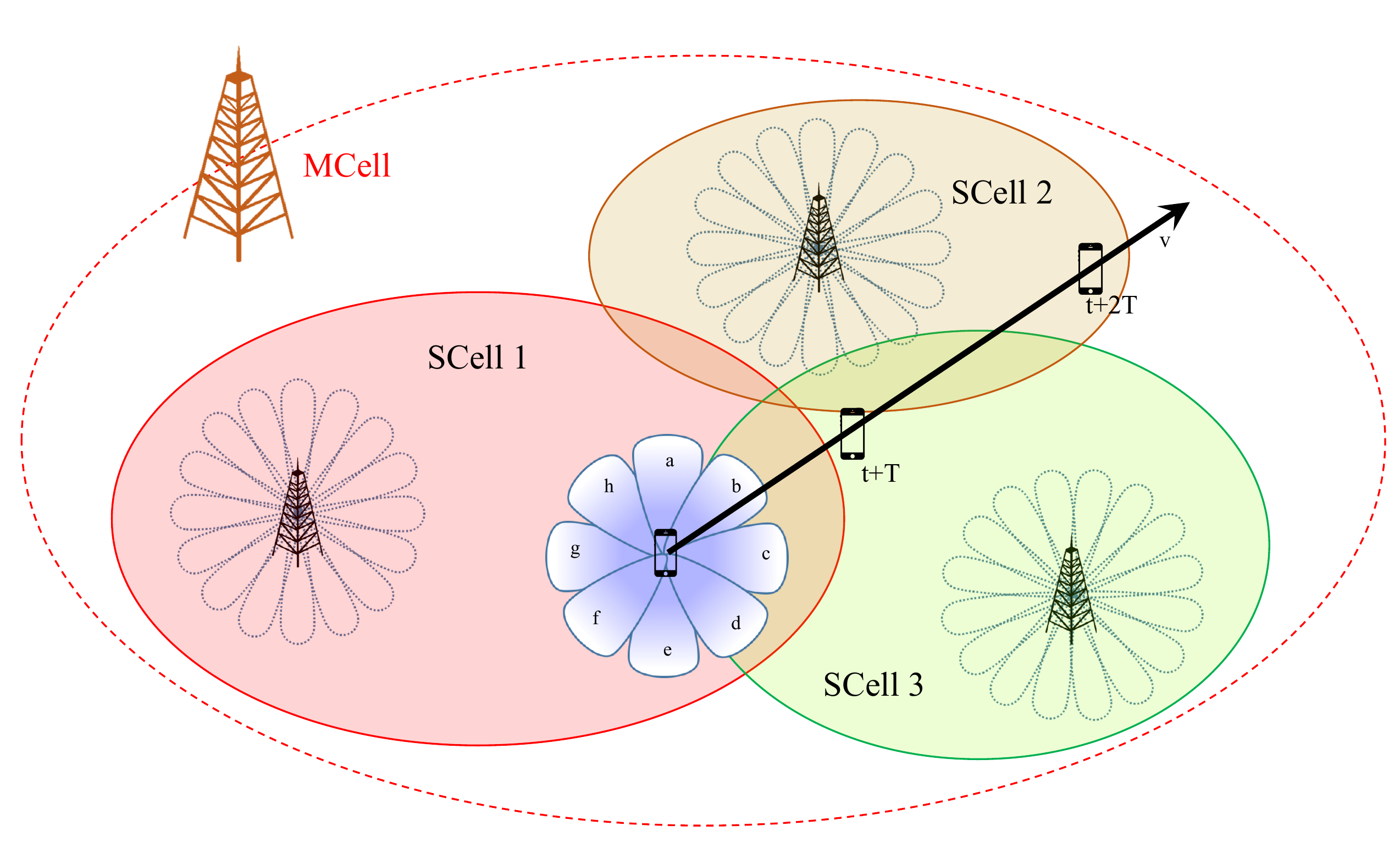}
 \caption{Handover in mmWave cellular networks. UE moves at constant velocity $v$. Solid (and dashed) ellipses show mmWave (and microwave) coverage boundaries (idealized for ease of  discussion).}
 \vspace{-0.6cm}
 \label{fig:HO}
\end{figure}

Our proposed procedure exploits the centralized-MCell control over the network, which can be used to determine the UE optimal target mmWave SCell (and direction) to associate with, as illustrated in Figure \ref{fig:HO}, when the user is in \emph{connected-mode}, i.e., it is already synchronized with both the macro and the mmWave cells. The key input information for the handover decision includes (i) instantaneous channel quality, (ii)  channel variance, and (iii) cell occupancy.

Assume that a new RT is collected at the SCell side and forwarded to the MCell, after the proposed three-phase MC procedure has been completed. 
By accessing the table, if the optimal mmWave cell the user should associate to is different from the current one, then a handover should be triggered, to maximize the user SINR and rate\footnote{In order to reduce the handover frequency, more sophisticated decision criteria could be investigated, rather than triggering a handover every time a more suitable SCell is identified (i.e., the reassociation might be performed only if the SINR increases above a predefined threshold, with respect to the previous time instant). A more detailed discussion of the different handover paradigms is beyond the scope of this paper.}. 
The use of both microwave and mmWave control planes is a key functionality for such an handover technique. 
In fact, in Phase 3, the handover decision is forwarded to the UE by the MCell, whose microwave link is much more robust and less volatile than its mmWave counterpart, thereby removing a point of failure in the control signaling path. Since each SCell periodically forwards the RT, the MCell has a complete overview of the cell dynamics and propagation conditions and can accordingly make handover decisions, to maximize the overall performance of the cell it oversees.

As an example, we refer to Figure \ref{fig:HO}. Suppose that, at time $t$, the user is attached to SCell 1 and keeps moving in a fixed direction at constant speed $v$. When, at time $t + T$, the MCell collects a new CRT, it can decide whether or not to make the user perform a handover to another mmWave cell (i.e., SCell 3, if the highest SINR is no longer referred to SCell 1, meaning that the scenario has changed and a handover to another mmWave cell could increase the user's QoS).

We finally remark that if previous versions of the report table are kept as a record, the  MCell can also use the SCells variance in selecting the mmWave cell a user should attach to, after a handover is triggered. If a selected SCell shows a large variance (which reflects high channel instability), the user might need to handover again in the very near future. Therefore, it could be better to trigger a handover only to an SCell which grants both good SINR and sufficient channel stability, leading to a more continuous and longer-term association with such designated cell.

\subsection{Multi-Cell Initial Access}
\label{sec:MC-IA}

\begin{figure}[t!]
\centering
 \includegraphics[trim= 0cm 0cm 0cm 0cm , clip=true, width=1 \columnwidth]{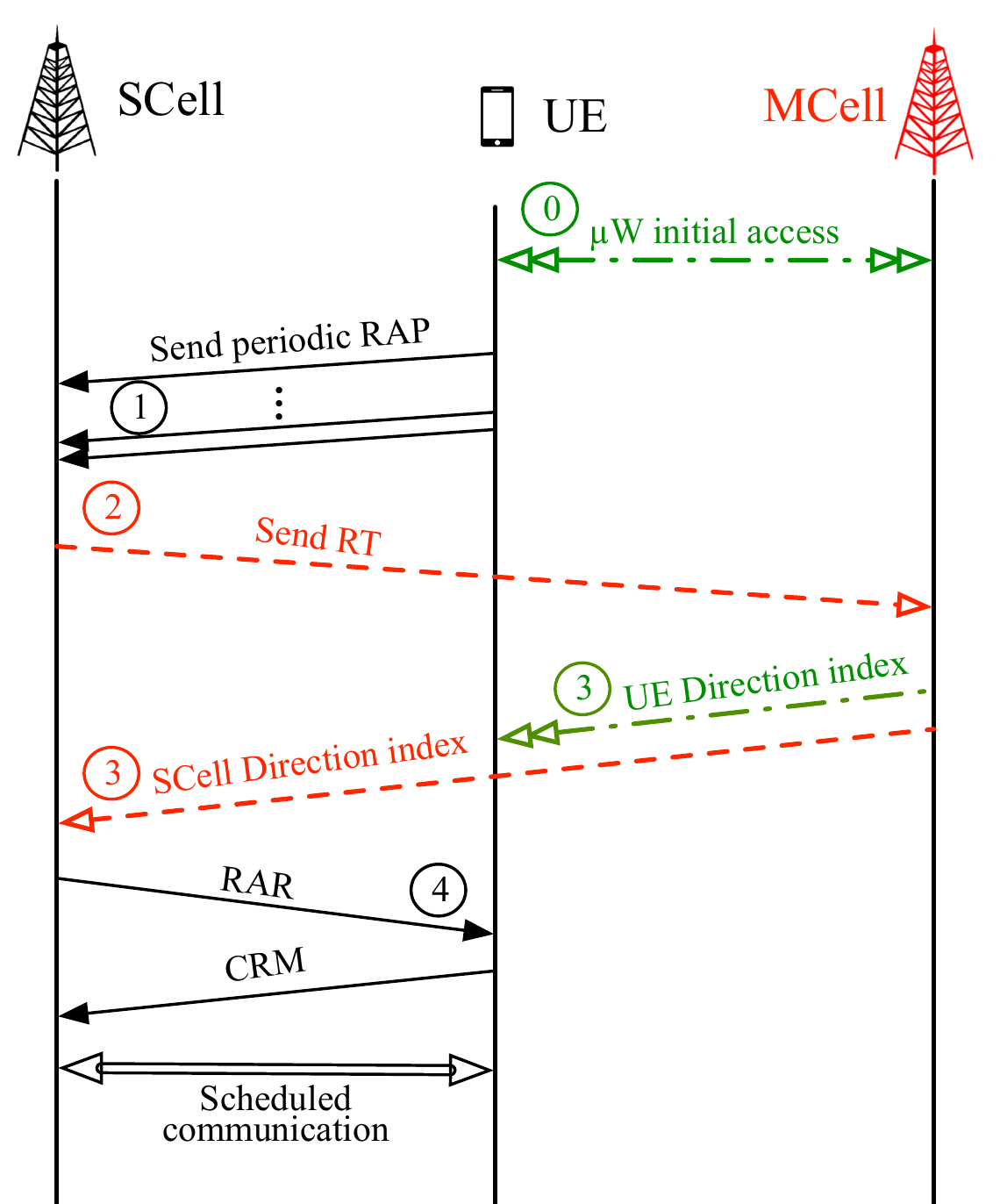}
 \caption{Proposed multicell initial access procedure. Both the macro MCell and the SCell  base stations play a role in this cell selection technique, where the presented MC algorithm is also used. Red and green dashed lines refer to the control messages exchanged via the bachaul X2 and the legacy communication links, respectively.}
 \label{fig:initial_access_procedure}
\end{figure}

The  procedure described in Sections \ref{sec:MCP} and \ref{sec:HO} referred to a UE that is already connected to the network.
However, we show that the uplink based control may be leveraged for fast initial access
from idle mode.
Initial access (IA) is the procedure by which a mobile UE establishes an initial 
physical link connection with a cell, a necessary step to access the network.
In current LTE systems, IA is performed on omnidirectional channels \cite{LTE_book}. 
However, in mmWave cellular systems, transmissions will need to be directional 
to overcome the increased isotropic pathloss experienced at higher frequencies.
IA must thus 
provide a mechanism by which the BS and the UE can determine suitable initial directions of transmission.

MmWave IA  procedures have been recently analyzed in \cite{Barati,barati2015initial,Capone,alk}. 
Different design options have been compared in \cite{CISS} and \cite{magazine_IA}, to evaluate coverage and access delay. We refer to \cite{magazine_IA} for a more detailed survey of recent IA works.
All of these methods are based on the current LTE design where each cell broadcasts
synchronization signals and the UE scans the directional space to detect the signals
and base station cells to potentially connect to.
We will call these ``downlink" based designs, since the transmissions come from the 
BSs to the UEs.
A key result of these findings is that the dominant delay in downlink-based IA arises in this initial 
sychronization phase.

In this work, we propose an alternate ``uplink"  scheme, mainly based on the proposed MC procedure described in Section \ref{sec:MCP},
as shown in Figure \ref{fig:initial_access_procedure}.  A UE first searches for
synchronization signals from conventional 4G cells.  This detection is fast since it 
can be performed omni-directionally and there is no directional scanning.  
Now, under the assumption that the
5G mmWave cells are roughly time synchronized to the 4G cell, and the round trip propagation times
are not large, an uplink transmission from the UE will be roughly time aligned at any 
closeby mmWave cell.  For example, if the cell radius is 150~m (a typical mmWave cell),
the round trip delay is only 3~$\mu$s.  A UE desiring initial access
thus broadcasts a random access preamble (RAP) scanning different angular directions.
Each of these RAPs will arrive roughly time-aligned in the random access slots of
all potential neighboring mmWave cells.  The mmWave cells will scan for the presence  of RA preambles, and when the RA preambles are detected, the MCell  performs the best attachment decision, based on the received RTs, feeding back the choice to the UE through the 4G-LTE link.

To compare uplink and downlink based IA, first suppose that the BS and the UE
can transmit and receive in only one direction at a time.  In this case, in either
DL or UL-based synchronization, the BS and the UE must search all $N_{\rm BS}N_{\rm UE}$ 
direction pairs and hence both UL and DL-based IA will take roughly the same time.
To reduce the search time, the receiver (the BS in the UL case
and the UE in the DL case) must be able to search in multiple directions simultaneously,
via either hybrid or digital beamforming.  Suppose that the receiver can 
look in $L$ directions.  Then, the scan time would be reduced by a factor
of $L$ to $N_{\rm BS}N_{\rm UE}/L$.  In particular, if the BS could perform fully digital reception
and hence look in all $L=N_{\rm BS}$ directions at the same time, the UL-based IA would require
only $N_{\rm UE}$ scans.  Similarly, if the UE could perform fully digital reception, 
the DL procedure would require $N_{\rm BS}$ scans.  

The reason the uplink-based IA may be preferable is that hybrid or fully digital 
receivers are more costly in terms of power consumption, and hence are more likely to be implemented
in a BS rather than in a UE.  In this case, the delay gains can be significant.
We will evaluate these gains precisely in Section \ref{sec:delay}.

\section{Simulation Model and Results}
\label{sec:res}

In this section, we present simulations to show that
(i) Multi-cell IA can offer significantly reduced latency
in the presence of digital beamforming at the BS; and 
(ii) at reasonable cell densities, a UE can see multiple mmWave cells with high probability.
This latter point suggests that multiple connectivity can have significant benefits in practical systems,
although further capacity evaluation will be needed.

\subsection{Simulation Parameters}

The parameters that we use to run our simulations are based on realistic system design considerations and are summarized in Table \ref{tab:params}.
 \renewcommand{\arraystretch}{1.3}
\begin{table}[!t]
\centering
\begin{tabular}{|c|c|c|}
\hline
\textbf{Parameter} & \textbf{Value} & \textbf{Description}\\
\toprule
\hline
 $W_{\rm tot}$ & $1$ GHz & Total system bandwidth\\
\hline
DL  $P_{\rm TX}$ & $30$ dBm & Transmission power \\
\hline
 NF  & $5$ dB & Noise figure \\
\hline
$f_{\rm c}$ & $28$ GHz & Carrier frequency \\
\hline
$\tau$ & $ -5$ dB &  Minimum SNR threshold \\
\hline
  SCell antenna & $8 \times 8$  & BS UPA MIMO array size  \\
\hline
UE antenna & $4 \times 4$ & UE UPA MIMO array size\\
\hline
$N_{\rm BS}$& $16$  & SCells scanning directions  \\
\hline
$N_{\rm UE}$& $8$  & UE scanning directions  \\
\hline
 $\lambda_{\rm BS}$ & varied & BS density per km$^2$ \\
\hline
 A & $0.5$ km$^2$ & Area of the simulation \\
\hline
$T_{\rm sig}$ & $10 \, \: \mu s$& Signal duration \\
\hline
$\phi_{\rm ov}$ & $5\%$ & Overhead\\
\hline
$T_{\rm per}$ & $200 \: \mu$s & Period between transmissions \\
\hline
\end{tabular}
\caption{Simulation parameters.}
\label{tab:params}
\end{table}

The channel model we have implemented is based on recent real-world measurements at $28$ GHz in New York City, to provide a realistic assessment of mmWave micro and picocellular networks in a dense urban deployment. Statistical models are derived for key channel parameters, including: (i) a distance-based pathloss, which models line-of-sight (LOS), non-line-of-sight (NLOS) and outage conditions; (ii) spatial clusters, described by central azimuth and elevation angles, fractions of power and angular beamspreads;  (iii) a small-scale fading model, where each of the path clusters is synthesized with a large number of subpaths, each  having its own peculiarities on horizontal and vertical angles (generated around the cluster central angles). Further details on the channel model and its parameters can be found in \cite{Mustafa,ns3_nokia,rappaport_channel_model}.

Our results are derived through a Monte Carlo approach, where multiple independent simulations are repeated, to  get different statistical quantities of interest. In each experiment: (i) we deploy multiple mmWave base stations and a UE, according to a Poisson Point Process (PPP), as done in \cite{Heath}; (ii) we perform the multi-connectivity algorithm by establishing a mmWave link between each SCell-UE pair and collecting the SINR values that each SCell perceives, when the transceiver performs the sequential scan; and (iii) we select the most profitable mmWave cell the user should attach to, according to the maximum saved SINR entry.

%The SINR between a BS $\rm j$ and UE is computed in the following way:
%\begin{equation}
%\text{SINR}_{\rm j,UE} = \frac{\frac{P_{\rm TX}}{PL_{\rm j,UE}}G_{\rm j,UE}}{\sum_{\rm k\neq j}\frac{P_{\rm TX}}{PL_{\rm k,UE}}G_{\rm k,UE}+W_{\rm tot}\times N_0}
%\label{eq:SINR}
%\end{equation}
%where $PL$ represents the pathloss, $G$ accounts for the beamforming gain that can be achieved when the transceiver exploits a MIMO antenna array, $P_{\rm TX}$ is the transmission power and $ W_{\rm tot}\times N_0$ is the thermal noise.
%In \eqref{eq:SINR}, it is assumed that the UE is interfered by $k$ other transmitters. However, to some extent, all cells have their own Control-plane (C-plane) which will removes any inter-cellular interference. According to the different instantaneous SCell and UE pointing directions, different beamforming gains are obtained and potentially different SINR values are collected and stored in the respective RT cells.

Referring explicitly to the MC procedures, we will consider an SINR threshold $\tau = -5$ dB, assuming that, if $\text{SINR}_{i,j}(k) < \tau$,  the SCell does not collect any control signal when the UE  transmits through direction $i$  and the BS $j$ is steering through direction $k$. 
Reducing  $\tau$ allows the user to be potentially found by more suitable mmWave cells, at the cost of designing more complex (and expensive) receiving schemes, able to detect the intended signal in more noisy channels. A set of two dimensional antenna arrays is used at both the mmWave SCells and the UE. BSs are equipped with a Uniform Planar Array (UPA) of $8 \times 8$ elements, which allow them to steer beams in $N_{\rm BS}=16$ directions; the user exploits an array of $4 \times 4$ antennas, steering beams through $N_{\rm UE}=8$ angular directions. The spacing of the elements is set to $\lambda/2$, where $\lambda$ is the wavelength.  
According to \cite{Barati} and \cite{CISS}, we assume that the SRSs  are transmitted periodically once every $T_{\rm per} = 200 \: \mu s$, for a duration of $T_{\rm sig}=10 \: \mu s$ (which is deemed to be sufficient to allow proper channel estimation at the receiver), to maintain a constant overhead of $\phi_{\rm ov} = T_{\rm sig}/T_{\rm per} = 5 \%$.

%Since, in the first phase, the UE sends $N_{\rm UE} \cdot N_{\rm BS} = 16 \cdot 8 = 128$ SRSs, each SCell will build a new RT and each macro MeNB will have a new channel conditions update every $(T_{\rm per}+T_{\rm sig})\cdot 128 = 26.88$ ms (when using analog BF at the transceiver)\footnote{We should also take care of the latency introduced by the feedback that the MeNB sends to the user and the mmWave SCells in the considered area, to perform the attachment procedure.}.

\subsection{Delay with Multi-Cell IA}
\label{sec:delay}

We first assess the delay with Multi-Cell IA in Section~\ref{sec:MC-IA}.
Following \cite{Barati,barati2015initial},
we suppose that in either the uplink or the downlink
direction, the random access or synchronization signals are $T_{\rm sig}$ long
and occur once ever $T_{\rm per}$ seconds.  The size of $T_{\rm sig}$ is
determined by the necessary link budget and we will assume that it is the same in either
direction.  The values in Table~\ref{tab:params} are based on simulations in 
\cite{barati2015initial} that enable reliable detection with an overhead of $T_{\rm sig}/T_{\rm per}$
of 5\%.  Now, as discussed in Section~\ref{sec:MC-IA}, the scanning for either the synchronization 
signal in the downlink or the random access preamble in the uplink will require
$N_{\rm BS}N_{\rm UE}/L$ scans, where $L$ is the number of directions that the 
receiver can look in at any one time.  Since there is one scanning opportunity every $T_{\rm per}$
seconds, the total delay is
\[
    D = \frac{N_{\rm BS}N_{\rm UE}T_{\rm per}}{L}.
\]
The value of $L$ depends on the beamforming capabilities.  In the uplink-based design, $L=1$ if
the BS receiver has analog BF and $L=N_{\rm BS}$ if it has a fully digital transceiver.
Similarly, in the downlink
$L=1$ if the UE receiver has analog BF and $L=N_{\rm UE}$ if it has a fully digital transceiver.
Table \ref{tab:BF_arch} compares the resulting delays for UL and DL-based designs
depending on the digital BF capabilities of the UE and the BS.  As discussed above,
digital BF is much more likely at the BS than at the UE due to power requirements.
We see that, in this case, the UL design offers significantly reduced access delay.

\begin{table}[h!]
\centering
\renewcommand{\arraystretch}{1.2}% Tighter
\begin{tabularx}{1\columnwidth}{@{\extracolsep{\fill}}c |c| c|c}
\hline
\multicolumn{2}{c|}{\textbf{BF Architecture}} & \multicolumn{1}{c|}{\begin{tabular}{@{}c @{}} \medskip \quad \\ \textbf{DL-based} \\ SCell transmits \\ UE receives   \end{tabular}} & \multicolumn{1}{c}{\begin{tabular}{@{}c @{}} \medskip \quad \\ \textbf{UL-based} \\  SCell receives \\ UE transmits  \end{tabular}} \\
\cline{1-2}
\textbf{ SCell Side} & \textbf{UE Side} & & \\
 \hline  \hline
 Analog & Analog & $N_{\rm UE} N_{\rm BS} $ ($25.6$ ms) & $N_{\rm UE} N_{\rm BS}$ ($25.6$ ms)  \\
  \hline
 Analog & Digital & $N_{\rm BS} $ ($3.2$ ms)  & $N_{\rm UE} N_{\rm BS} $ ($25.6$ ms)\\
  \hline
 Digital & Analog & $N_{\rm UE} N_{\rm BS} $ ($25.6$ ms)& $N_{\rm UE} $ ($1.6$ ms)\\
  \hline
 Digital & Digital & $N_{\rm BS} $ ($3.2$ ms) & $ N_{\rm UE} $ ($1.6$ ms)\\
\hline
\end{tabularx}
\caption{Number of  synchronization signals (or RAPs) that the BS (or the UE) has to send (and corresponding time) to perform a DL (or UL) based procedure. A comparison among different BF architectures (analog and fully digital) is performed. We assume $T_{\rm sig} = 10 \: \mu$s, $T_{\rm per} = 200 \: \mu$s (to maintain an overhead $\phi_{\rm ov} = 5\%$), $N_{\rm UE} = 8$ and $N_{\rm BS} = 16$.}
\label{tab:BF_arch}
\end{table}

\subsection{Number of Cells}

\begin{figure}[t!]
\centering
 \includegraphics[trim= 0cm 0cm 0cm 0cm , clip=true, width=0.95\columnwidth]{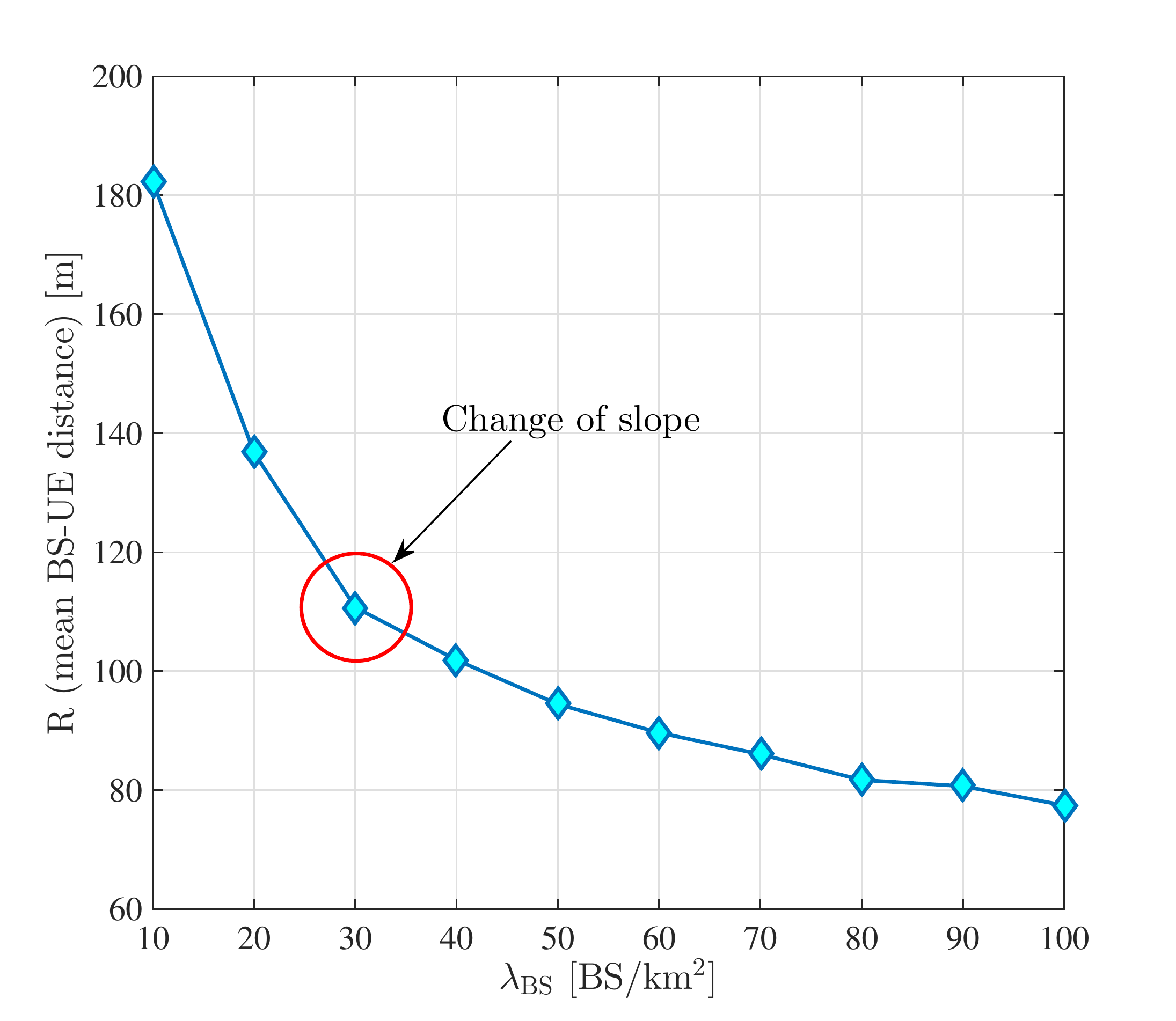}
 \caption{Mean SCell-UE distance (R) vs. SCell density ($\lambda_{\rm BS}$). A change of slope is identified when $\lambda_{\rm BS}>30$ BS/km$^2$.}
 \label{fig:D_R}
\end{figure}

We next try to assess the number of cells that a UE can typically see.
In Figure  \ref{fig:D_R} we depict the mean distance $R$ of a user to its  serving mmWave cell versus the SCell density $\lambda_{\rm BS}$ in the selected area. The results show that, when the number of active SCells increases, the  user can more likely find a closer cell to attach to. In this way, its propagation conditions become less demanding and the UE can reliably achieve a higher throughput. We note that a change of slope occurs when $\lambda_{\rm BS}>30$ BS/km$^2$, reflecting in a slower reduction of $R$. In fact the initial sudden drop of $R$, when increasing the BS density, reflects the transition from a user outage regime to a LOS/NLOS regime \cite{Mustafa}.
 After that,  as we  persistently keep  densifying the network, the SCells become so crowded that the user hardly  finds a cell to build a good mmWave link with, regardless of the actual density.
  When this steady state is reached, the deployment of more SCells leads to a considerable increase of the system complexity, while providing a limited reduction of $R$.

\begin{figure}[t!]
\centering
 \includegraphics[trim= 0cm 0cm 0cm 0cm , clip=true, width=0.95 \columnwidth]{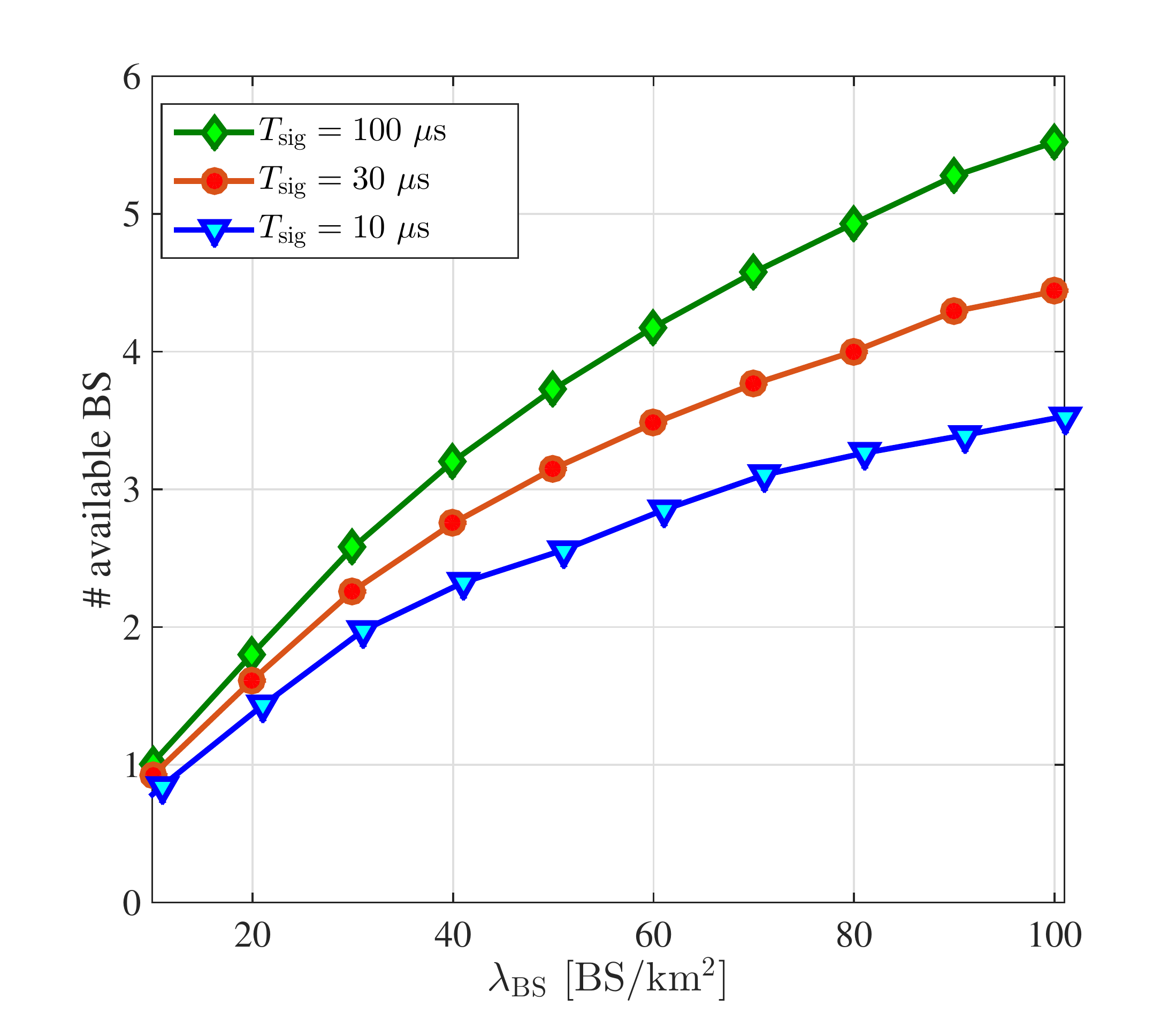}
 \caption{Average number of active and potentially available SCells (capable of granting SINR above threshold for the considered user) versus the SCell density $\lambda_{\rm BS}$. Different control signal durations $T_{\rm sig}$ are considered.}
 \label{fig:BS_vs_D}
\end{figure}

We finally show, in Figure  \ref{fig:BS_vs_D}, the average number of active and potentially available BSs (capable of granting SINR above threshold for the considered user) versus the SCell density $\lambda_{\rm BS}$, when different control signal durations $T_{\rm sig}$ are considered.
We see that, when increasing the mmWave cell density, the set of cells that could possibly serve the user is increased. After  Phase 2 of our proposed procedure, the macro MCell will select just one SCell for the user (according to the maximum perceived SINR). However, having at least a second active BS  to which the user may connect, if its primary cell's link is partially unavailable (e.g., due to a blockage event), can add a robustness to the network and ensure a sufficiently  good QoS. From Figure \ref{fig:BS_vs_D}, we see that there should be at least $\lambda_{\rm BS}=30$ BS/km$^2$, when $T_{\rm sig} = 10 \: \mu$s, in order to have, on average, at least $2$ available BSs.
The number of available SCells can be increased also by increasing the signal duration: if $T_{\rm sig}$ is increased, each UE transmits its SRSs for a longer time in the same sector, so that each SCell belonging to that sector can accumulate a higher amount of energy, which results in an increased SINR.
%In such a way, if the signal duration is doubled, the SINR also doubles and this is equal to reducing $\tau$ by $3$ dB.
As an example, by considering $T_{\rm sig}=100 \: \mu$s, we just need $\lambda_{\rm BS}=20$ BS/km$^2$ to have, on average, at least $2$ available SCells. This will of course increase the overall duration of the  procedure, leading to latency and delay drawbacks.

\section{Conclusions and Future Work}
\label{conclu}

A  challenge for the feasibility of a 5G mmWave system is the high susceptibility to blocking that affects links in mmWave networks and results in rapid channel dynamics.
In order to deal with these channel variations, a periodical directional sweep should be performed, to constantly monitor the directions of transmission of each potential link and to adapt the beam steering when a power signal drop is detected.
In this work, we have proposed a novel measurement reporting system that allows a supervising centralized entity, such as the macro base station, operating in the legacy band, to periodically collect multiple reports on the overall channel propagation conditions, that can be used to make proper network decisions when implementing multiple control-plane features, such as initial access or handover.
We argue that this proposed approach, which is based on uplink rather than downlink signals, can enable much more rapid and robust tracking, enabling also the use of digital beamforming architectures to dramatically reduce the measurement reporting delay.

As part of our future work, we aim at analyzing in greater detail the proposed control-plane applications that our reporting technique is suitable for, showing that keeping a record of the received signal strength variance can highly benefit the user and improve the overall network control performance. Moreover, a study on the implementation of analog, hybrid or fully digital beamforming architectures for such control-plane applications and a comparison among them deserves a deeper investigation.

\bibliographystyle{IEEEtran}
\bibliography{bibliography}

\begin{thebibliography}{10}
\providecommand{\url}[1]{#1}
\csname url@samestyle\endcsname
\providecommand{\newblock}{\relax}
\providecommand{\bibinfo}[2]{#2}
\providecommand{\BIBentrySTDinterwordspacing}{\spaceskip=0pt\relax}
\providecommand{\BIBentryALTinterwordstretchfactor}{4}
\providecommand{\BIBentryALTinterwordspacing}{\spaceskip=\fontdimen2\font plus
\BIBentryALTinterwordstretchfactor\fontdimen3\font minus
  \fontdimen4\font\relax}
\providecommand{\BIBforeignlanguage}[2]{{%
\expandafter\ifx\csname l@#1\endcsname\relax
\typeout{** WARNING: IEEEtran.bst: No hyphenation pattern has been}%
\typeout{** loaded for the language `#1'. Using the pattern for}%
\typeout{** the default language instead.}%
\else
\language=\csname l@#1\endcsname
\fi
#2}}
\providecommand{\BIBdecl}{\relax}
\BIBdecl

\bibitem{RanRapE:14}
S.~Rangan, T.~S. Rappaport, and E.~Erkip, ``Millimeter-wave cellular wireless
  networks: Potentials and challenges,'' \emph{Proceedings of the IEEE}, vol.
  102, no.~3, pp. 366--385, March 2014.

\bibitem{Mustafa}
M.~R. Akdeniz, Y.~Liu, M.~K. Samimi, S.~Sun, S.~Rangan, T.~S. Rappaport, and
  E.~Erkip, ``Millimeter wave channel modeling and cellular capacity
  evaluation,'' \emph{IEEE Journal on Selected Areas in Communications},
  vol.~32, no.~6, pp. 1164--1179, June 2014.

\bibitem{MC}
F.~B. Tesema, A.~Awada, I.~Viering, M.~Simsek, and G.~P. Fettweis, ``Mobility
  modeling and performance evaluation of multi-connectivity in 5{G}
  intra-frequency networks,'' in \emph{2015 IEEE Globecom Workshops (GC
  Wkshps)}, Dec 2015.

\bibitem{lu2012modeling}
J.~Lu, D.~Steinbach, P.~Cabrol, and P.~Pietraski, ``Modeling the impact of
  human blockers in millimeter wave radio links,'' \emph{ZTE Commun. Mag},
  vol.~10, no.~4, pp. 23--28, 2012.

\bibitem{CA_LTE}
M.~Iwamura, K.~Etemad, M.~h.~Fong, R.~Nory, and R.~Love, ``Carrier aggregation
  framework in 3{G}{P}{P} {L}{T}{E}-advanced,'' \emph{IEEE Communications
  Magazine}, vol.~48, no.~8, pp. 60--67, August 2010.

\bibitem{LTE_book}
S.~Sesia, I.~Toufik, and M.~Baker, \emph{LTE, The UMTS Long Term Evolution:
  From Theory to Practice}.\hskip 1em plus 0.5em minus 0.4em\relax Wiley
  Publishing, 2009.

\bibitem{zorzi}
H.~Shokri-Ghadikolaei, C.~Fischione, G.~Fodor, P.~Popovski, and M.~Zorzi,
  ``Millimeter wave cellular networks: A {M}{A}{C} layer perspective,''
  \emph{IEEE Transactions on Communications}, vol.~63, no.~10, pp. 3437--3458,
  Oct 2015.

\bibitem{handoff}
A.~Talukdar, M.~Cudak, and A.~Ghosh, ``Handoff rates for millimeterwave 5{G}
  systems,'' in \emph{IEEE 79th Vehicular Technology Conference (VTC Spring)},
  May 2014.

\bibitem{HO_train}
H.~Song, X.~Fang, and L.~Yan, ``Handover scheme for 5{G} {C}/{U} plane split
  heterogeneous network in high-speed railway,'' \emph{IEEE Transactions on
  Vehicular Technology}, vol.~63, no.~9, pp. 4633--4646, Nov 2014.

\bibitem{HO_het}
S.~Sadr and R.~S. Adve, ``Handoff rate and coverage analysis in multi-tier
  heterogeneous networks,'' \emph{IEEE Transactions on Wireless
  Communications}, vol.~14, no.~5, pp. 2626--2638, May 2015.

\bibitem{HO_het_2}
P.~Coucheney, E.~Hyon, and J.~M. Kelif, ``Mobile association problem in
  heterogenous wireless networks with mobility,'' in \emph{IEEE 24th Annual
  International Symposium on Personal, Indoor, and Mobile Radio Communications
  (PIMRC)}, Sept 2013, pp. 3129--3133.

\bibitem{Barati}
C.~N. Barati, S.~A. Hosseini, S.~Rangan, P.~Liu, T.~Korakis, S.~S. Panwar, and
  T.~S. Rappaport, ``Directional cell discovery in millimeter wave cellular
  networks,'' \emph{IEEE Transactions on Wireless Communications}, vol.~14,
  no.~12, pp. 6664--6678, Dec 2015.

\bibitem{barati2015initial}
\BIBentryALTinterwordspacing
C.~N. Barati, S.~A. Hosseini, M.~Mezzavilla, P.~Amiri-Eliasi, S.~Rangan,
  T.~Korakis, S.~S. Panwar, and M.~Zorzi, ``Directional initial access for
  millimeter wave cellular systems,'' in \emph{49th Asilomar Conference on
  Signals, Systems and Computers}, Nov 2015, pp. 307--311, \textit{CoRR,} vol.
  abs/1511.06483. [Online]. Available: \url{http://arxiv.org/abs/1511.06483}
\BIBentrySTDinterwordspacing

\bibitem{Capone}
A.~Capone, I.~Filippini, and V.~Sciancalepore, ``Context information for fast
  cell discovery in mm-{W}ave 5{G} networks,'' in \emph{Proceedings of 21th
  European Wireless Conference}, May 2015.

\bibitem{alk}
V.~Desai, L.~Krzymien, P.~Sartori, W.~Xiao, A.~Soong, and A.~Alkhateeb,
  ``Initial beamforming for mm{W}ave communications,'' in \emph{48th Asilomar
  Conference on Signals, Systems and Computers}, Nov 2014, pp. 1926--1930.

\bibitem{CISS}
M.~Giordani, M.~Mezzavilla, C.~N. {Barati Nt.}, S.~Rangan, and M.~Zorzi,
  ``Comparative analysis of initial access techniques in {5G} mmwave cellular
  networks,'' in \emph{Annual Conference on Information Science and Systems
  (CISS)}, Princeton, USA, 2016.

\bibitem{magazine_IA}
\BIBentryALTinterwordspacing
M.~Giordani, M.~Mezzavilla, and M.~Zorzi, ``Initial access in 5{G} mm-{W}ave
  cellular networks,'' \emph{CoRR}, vol. abs/1602.07731, 2016. [Online].
  Available: \url{http://arxiv.org/abs/1602.07731}
\BIBentrySTDinterwordspacing

\bibitem{ns3_nokia}
M.~K. Samimi and T.~S. Rappaport, ``{3-D} statistical channel model for
  millimeter-wave outdoor mobile broadband communications,'' in \emph{Proc.\
  ICC}, June 2015, pp. 2430--2436.

\bibitem{rappaport_channel_model}
Y.~Azar, G.~N. Wong, K.~Wang, R.~Mayzus, J.~K. Schulz, H.~Zhao, F.~Gutierrez,
  D.~Hwang, and T.~S. Rappaport, ``28 {G}{H}z propagation measurements for
  outdoor cellular communications using steerable beam antennas in {N}ew {Y}ork
  city,'' in \emph{IEEE International Conference on Communications (ICC)}, June
  2013, pp. 5143--5147.

\bibitem{Heath}
T.~Bai and R.~W. Heath, ``Coverage and rate analysis for millimeter-wave
  cellular networks,'' \emph{IEEE Transactions on Wireless Communications},
  vol.~14, no.~2, pp. 1100--1114, Feb 2015.

\end{thebibliography}

\end{document}